\documentclass[aps,prl,amsmath,amssymb,showpacs,twocolumn,superscriptaddress]{revtex4-1}

\usepackage{graphicx} 
\usepackage{subfigure} 

\def\k{{\bf k}} 
 
\def\q{{\bf q}}

\begin{document}

\newcommand{\etal}{{\it et al.}\/} 
\newcommand{\gtwid}{\mathrel{\raise.3ex\hbox{$>$\kern-.75em\lower1ex\hbox{$\sim$}}}} 
\newcommand{\ltwid}{\mathrel{\raise.3ex\hbox{$<$\kern-.75em\lower1ex\hbox{$\sim$}}}} \setcounter{footnote}{0} 
\renewcommand{\thefootnote}{\fnsymbol{footnote}}

\title{Inelastic neutron and x-ray scattering as probes of the sign structure of the Fe-pnictide superconducting gap}

\author{T. A. Maier} \affiliation{Center for Nanophase Materials Sciences and Computer Science and Mathematics Division, Oak Ridge National Laboratory, Oak Ridge, TN 37831-6494}

\author{S. Graser} \affiliation{Center for Electronic Correlations and Magnetism, Institute of Physics, University of Augsburg, D-86135 Augsburg, Germany}

\author{P. J. Hirschfeld} \affiliation{Department of Physics, University of Florida, Gainesville, FL 32611, U.S.A.}

\author{D. J. Scalapino} \affiliation{Department of Physics, University of California, Santa Barbara, CA 93106-9530 USA}

\date{\today}
\begin{abstract}
	Neutron spin-flip scattering observations of a resonance in the superconducting state is often taken as evidence of an unconventional superconducting state in which the gap changes sign $\Delta(k+Q)=-\Delta(k)$ for momentum transfers $Q$ which play an important role in the pairing. Recently questions regarding this identification for the Fe-pnictide superconductors have been raised and it has been suggested that $\Delta(k+Q)=\Delta(k)$. Here we propose that inelastic neutron or x-ray scattering measurements of the spectral weight of a phonon of momentum $Q$ can distinguish between these two pairing scenarios. 
\end{abstract}

\maketitle

Determining the superconducting gap structure in the Fe-based superconductors has proven difficult due to the complicated multisheeted, multiorbital nature of the Fermi surface, as well as problems with fabricating clean films and interfaces. At present the most popular candidate for pairing in most of these materials is the $s^{+-}$ state which changes sign between hole and electron pockets and arises naturally from a spin-fluctuation interaction\cite{ref:Mazin,ref:Kuroki,ref:Graser}. However, there are few phase-sensitive tests of the relative sign change of $\Delta_\k$ between the different Fermi surfaces. Josephson tunneling experiments \cite{ref:Tsuei,ref:RGreene}, Fourier transform STM in a magnetic field \cite{ref:Hanaguri}, and the observation of a resonance mode in inelastic neutron scattering experiments\cite{ref:Christianson,ref:Lumsden,ref:Li,ref:Inosov} have been offered as limited evidence for such a sign change. 

Recently, an electron-phonon interaction enhanced by orbital fluctuations has been proposed as the mechanism which is responsible for pairing in the Fe-based superconductors \cite{ref:Kontani,ref:Saito,ref:Yanagi}. This interaction leads to a so-called $s^{++}$ gap which is positive on both the hole and electron Fermi surfaces. While in both cases, the gap can be anisotropic and accidental nodes can even occur, the relative sign of the gap on regions of the hole and electron Fermi surfaces which have the same dominant orbital weight will be negative if the pairing is driven by spin-fluctuations and positive if driven by phonon-orbital-fluctuations. Thus it is important to determine the relative sign between $\Delta(k)$ and $\Delta(k+Q)$ on the hole and electron Fermi surfaces, respectively, for momentum transfers $Q$ which connect such regions.

One test of this is the occurrence of a spin resonance peak in the neutron spin-flip scattering in the superconducting state \cite{ref:Monthoux}. The BCS coherence factor which enters the spin-flip scattering is 
\begin{equation}
	\frac{1}{2}\left(1-\frac{\Delta(k+Q)\Delta(k)}{E(k+Q)E(k)}\right). \label{eq:1} 
\end{equation}
Here $E(k)=\sqrt{\xi^2_k+\Delta^2(k)}$ is the quasi-particle energy and near threshold, where $\xi_k$ goes to zero, this coherence factor goes to 1 if $\Delta(k+Q)$ and $\Delta(k)$ have opposite signs and zero if they have the same sign. A resonance peak is expected in the former case \cite{ref:Maier,ref:Korshunov}. However, arguments \cite{ref:Onari} have been given that for the case of BaFe$_{1.85}$Co$_{0.15}$As$_2$ the experimentally observed peak \cite{ref:Christianson,ref:Lumsden,ref:Li,ref:Inosov} may be consistent with an $s^{++}$ gap, if the scattering rate collapses sufficiently rapidly in the superconducting state. Thus one would like to find additional experimental probes which provide a way of determining the relative sign of the gap between the hole and electron pockets.

As is known, there exist anomalies in the phonon spectral weight of conventional superconductors. Various authors \cite{ref:Zeyher,ref:Marsiglio,ref:Allen} have shown that if the phonon frequency $\Omega_q$ is greater than twice the gap $\Delta$, but not too far above, significant spectral weight is transferred below $2\Delta$ into a resonance. In addition, the phonon peak shifts to higher frequencies and broadens. These effects have been convincingly observed in neutron scattering experiments on the borocarbide superconductor YNi$_2$B$_2$C \cite{ref:Weber}. This behavior reflects the fact that in the superconducting state, the coherence factor that enters the coupling of a phonon to electron polarization fluctuations is similar to Eq.~(\ref{eq:1}) except that it has a plus sign. In this case the coherence factor will go to 1 at threshold if $\Delta(k+Q)$ and $\Delta(k)$ have the same sign. This will lead to an anomalous resonance and hardening of the phonon peak for an $s^{++}$ gap. Here we point out that for an $s^{+-}$ gap with $\Delta(k)=-\Delta(k+Q)$, the anomalous resonance is absent and the phonon peak shifts down (softens) slightly. Thus neutron and x-ray scattering \cite{ref:Reznik} measurements of the phonon spectral weight at specific $Q$ values can provide information on the relative sign of the gap \cite{ref:Flatte}.

\begin{figure}
	[htbp] 
    \includegraphics[width=7cm]{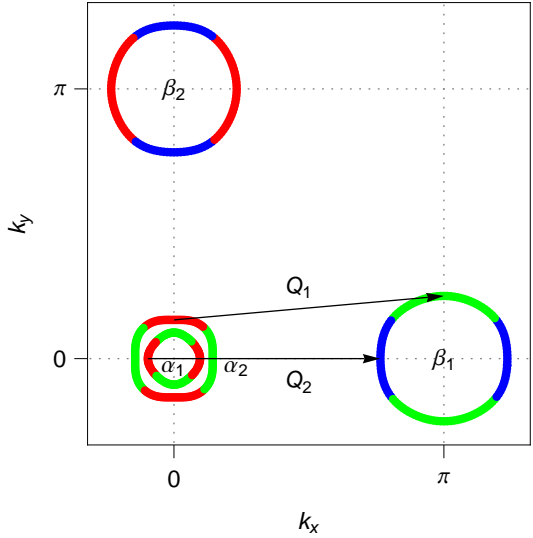} \caption{(Color online) Fermi surfaces for a five-orbital model of the Fe-pnictide superconductors. The colors indicate the dominant orbital character (red=$d_{xz}$, green=$d_{yz}$, blue=$d_{xy}$). Here $Q_1$ connects $d_{xz}$ parts of the hole $\alpha_2$ and $d_{yz}$ parts of the electron $\beta_1$ Fermi surfaces and $Q_2$ connects $d_{xz}$ parts of $\alpha_1$ to the $d_{xy}$ parts of $\beta_1$. 
	\label{fig:1}} 
\end{figure}

To explore the use of this effect to determine the relative gap signs in the Fe-based materials, we will use the same 5-orbital tight binding fit of the LDA bandstructure that we previously used to discuss the possible occurrence of a neutron scattering spin resonance \cite{ref:MGSH}. In this case, with an electron doping $x=0.125$ one finds the Fermi surfaces and orbital occupations illustrated in Fig.~\ref{fig:1}. Here there are two hole Fermi surfaces $\alpha_1$ and $\alpha_2$ around the $\Gamma$ point and two electron Fermi surfaces $\beta_1$ and $\beta_2$ around the $X(\pi,0)$ and $Y(0,\pi)$ points of the unfolded (1Fe/cell) Brillouin zone. We will use an orbital notation $\ell=1,2,3,4,5$ to denote the $d_{xz}$, $d_{yz}$, $d_{xy}$, $d_{x^2-y^2}$ and $d_{3z^2-r^2}$ orbits, respectively, where the axes are along the Fe-Fe bonds. The dominant orbital weight of the Bloch states on the Fermi surfaces is indicated in Fig.~\ref{fig:1}.

As Kontani and Onari \cite{ref:Kontani} have noted, local Fe-ion displacements couple to orbital fluctuations. The polarization susceptibility associated with these orbital fluctuations enters the phonon self-energy and is reflected in the phonon spectral weight. Here we examine the change in the phonon spectral weight $-\frac{1}{\pi} {\rm Im}\,D(Q,\omega)$ associated with the $Q_1=(\pi,0.09\pi)$ and $Q_2=(0.86\pi,0)$ momenta shown in Fig.~\ref{fig:1} when the system goes from a normal state to a superconducting state with a nominal $s^{++}$ gap or an $s^{+-}$ gap. We begin with $Q_1$ and will assume that there is an electron-phonon coupling $g_{12}$ for scattering an electron between the dominantly $d_{xz}$-like states on the $\alpha_2$ hole Fermi surface to the $d_{yz}$-like states on the $\beta_1$ electron Fermi surfaces. In the following, the real part of the normal state phonon self-energy will be absorbed to give the dressed normal state phonon frequency $\Omega_{Q_1}$. Then at low energies, the imaginary part of the phonon self-energy for momentum $Q_1$ dominantly comes from scattering between the nearly nested $d_{xz}$-like states on $\alpha_2$ and the $d_{yz}$-like states on $\beta_1$. Similarly, the change between the superconducting and normal state phonon self-energy is also determined by the $d_{xz}$--$d_{yz}$ orbital fluctuations. In this case the important orbital contribution to the phonon self-energy is 
\begin{equation}
	\Pi(q,\omega)=-|g_{12}|^2\left(\chi^{\rm orb}_{1212}(q,\omega)+\chi^{\rm orb}_{2121}(q,\omega)\right) \label{eq:2} 
\end{equation}
with 
\begin{equation}
	\chi^{\rm orb}_{1212}(q,\omega)=\left\{\chi^0(q,\omega)\left[1+U^c\chi^0(q,\omega)\right]^{-1}\right\}_{1212} \label{eq:3} 
\end{equation}
and similarly for $\chi^{\rm orb}_{2121}$.
Here the Coulomb interaction matrix is 
\begin{equation}
	U^c_{\ell_1\ell_2\ell_3\ell_4}=
	\begin{cases}
		\phantom{-}\bar U&\ell_1=\ell_2=\ell_3=\ell_4\\
		-\bar U'+2\bar J&\ell_1=\ell_3\ne\ell_2=\ell_4\\
		\phantom{-}\bar J&\ell_1=\ell_2\ne\ell_3=\ell_4\\
		\phantom{-}\bar J'&\ell_1=\ell_4\ne\ell_2=\ell_3\\
	\end{cases}
	\label{eq:4} 
\end{equation}
and 
\begin{eqnarray}
	\chi^0_{\ell_1\ell_2\ell_3\ell_4}(q)&=-\frac{T}{N}\sum_{k,\mu\nu} a^{\ell_1}_\mu(k+q)a^{\ell_3*}_\mu(k+q)a^{\ell_4*}_\nu(k)a^{\ell_3}_\nu(k)\cr &\times\bigl\{G^\mu(k+q)G^\nu(k)-F^\mu(-k-q)F^\nu(k)\bigr\}.\cr \label{eq:5} 
\end{eqnarray}
The notation follows Ref.~\cite{ref:Kemper} with $\bar U$ and $\bar U'$ the intra- and inter-orbital Coulomb interactions, $\bar J$ the exchange coupling and $\bar J'$ the pair hopping interactions. We will consider a typically rotationally invariant set of interaction parameters which are similar to those previously used to study the occurrence of a spin resonance in the $s^{+-}$ superconducting state \cite{ref:MGSH}. These are such that $\bar U=4$ in units of the hopping $t^{11}_y$, $\bar J=\bar U/16$, $\bar U'=\bar U-2\bar J$, and $\bar J'=\bar J$ \cite{Footnote1}. In the superconducting state the normal and anomalous Gor'kov Green's functions are given as 
\begin{equation}
	G^\mu(k)=\frac{i\omega_n+\xi_\mu(k)}{\omega^2_n+E^2_\mu(k)}\,,\,\, F^\mu(k)=\frac{\Delta_\mu(k)}{\omega^2_n+E^2_\mu(k)} \label{eq:6} 
\end{equation}
with $E_\mu=\sqrt{\xi_\mu^2(k)+\Delta_\mu^2(k)}$. For the normal state calculations, the gap is set to zero in Eq.~(\ref{eq:5}).

Here, because we are calculating the orbital-fluctuations, the sign between the normal and anomalous Green's function products in Eq.~(\ref{eq:5}) is negative. It would be positive if we were studying the spin susceptibility and in addition the orbital interaction matrix $U^c$ would be changed to $-U^s$. In Eq.~(\ref{eq:5}) we use a notation in which $k=(\k,\omega_n)$ and $q=(\q,(\omega_m\to\omega+i\delta))$. The calculations were carried out for a low temperature $T=0.05$ in units of $t_y^{11}$.
\begin{figure}
	[htbp] 
	\includegraphics[width=8cm]{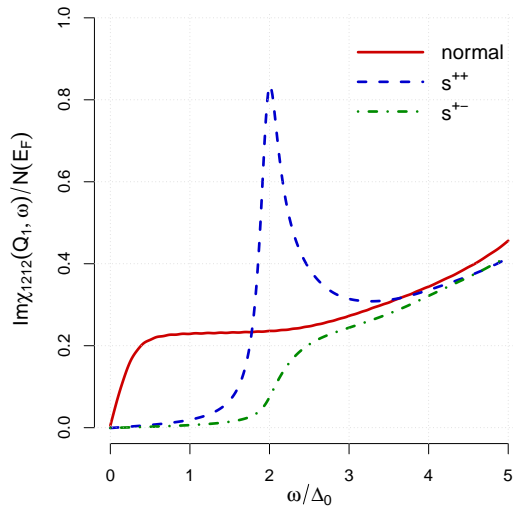} \caption{(Color online) The imaginary part of $\chi^{\rm orb}_{1212}(Q_1,\omega)$ versus $\omega$ for the normal state (red, solid), the $s^{++}$ superconducting state with $\Delta_0(k+Q_1)=\Delta_0(k)=\Delta_0$ (blue, dashed) and the $s^{+-}$ state with $\Delta_0(k+Q_1)=-\Delta_0(k)=\Delta_0$ (green, dot-dashed). \label{fig:2}} 
\end{figure}

Figure~\ref{fig:2} shows the imaginary part of $\chi^{\rm orb}_{1212}(q,\omega)$ versus $\omega$ with $q=Q_1$ for the normal state (solid), the superconducting states with an $s^{++}(\Delta_{\alpha_1}=\Delta_{\alpha_2}=\Delta_{\beta_1}=\Delta_{\beta_2}=\Delta_0)$ gap (dashed) and an $s^{+-}(\Delta_{\alpha_1}=\Delta_{\alpha_2}=-\Delta_{\beta_1}=-\Delta_{\beta_2}=\Delta_0)$ gap  (dash-dot) \cite{Footnote4}. The imaginary part of the orbital response is similar to what was found for the spin susceptibility with the $s^{+-}$ and $s^{++}$ gaps interchanged. Thus, if the phonon-orbital-fluctuations provide the dominant pairing interaction in the Fe-based superconductors then the gap will have $s^{++}$ character and one will find an anomalous resonance in the orbital susceptibility in the superconducting state rather than in the spin susceptibility.

Although it would be interesting to have a direct scattering probe of the orbital susceptibility spectral weight shown in Fig.~\ref{fig:2}, one can still access its structure by measuring the phonon spectral weight. Using the orbital susceptibility, we can evaluate the phonon self-energies that enter the normal and superconducting phonon progagators \cite{Footnote2},
\begin{equation}
	D^{-1}_N(q,\omega)=\frac{\omega^2-\Omega^2_{Q_1}}{2\Omega_{Q_1}}-i{\rm Im}\Pi_N(Q_1,\omega) \label{eq:7} 
\end{equation}
and 
\begin{equation}
	D^{-1}_S(q,\omega)=\frac{\omega^2-\Omega^2_{Q_1}}{2\Omega_{Q_1}}-{\rm Re}\delta\Pi(Q_1,\omega)-i{\rm Im}\Pi_S(Q_1,\omega). \label{eq:8} 
\end{equation}
Here $\delta\Pi=\Pi_S-\Pi_N$ and $\Pi_S$ and $\Pi_N$ are the superconducting and normal phonon self-energies obtained from Eq.~(\ref{eq:2}), with the gap set to zero in Eq.~(\ref{eq:5}) for the normal case and set to $\Delta(k)=\Delta(k+Q)=\Delta_0$ and $\Delta(k)=-\Delta(k+Q)=\Delta_0$ for the $s^{++}$ and $s^{+-}$ superconducting states, respectively.
\begin{figure}
	[htbp] 
	\includegraphics[width=8.0cm]{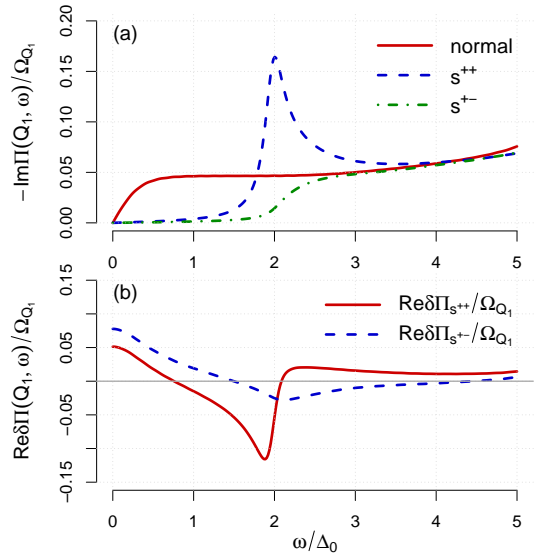} \caption{(Color online) (a) The imaginary part of the phonon damping $\Pi(Q_1,\omega)$ versus $\omega$ for the normal state (red, solid), the $s^{++}$ (blue, dashed) and $s^{+-}$ (green, dot-dashed) superconducting gaps. (b) The real part $\delta\Pi(Q_1,\omega)$ (see text) for the $s^{++}$ (red, solid) and $s^{+-}$ (blue, dashed) superconducting gaps. \label{fig:3}} 
\end{figure}

The phonon damping $-{\rm Im}\Pi(Q,\omega)/\Omega_{Q_1}$ versus $\omega/\Delta_0$ is plotted in Fig.~\ref{fig:3}a. Here we have taken $\Omega_{Q_1}=2.5\Delta_0$ and set $2|g_{12}|^2N(0)/\Omega_{Q_1}=0.2$. As $\Omega_{Q_1}$ approaches $2\Delta_0$, the effects we are discussing are enhanced and diminish when $\Omega_Q$ becomes large compared with $2\Delta_0$. As seen in Fig.~\ref{fig:3}a, for an $s^{++}$ gap there is an enhanced phonon damping above $2\Delta_0$. This arises from the increase in the density of states in this region due to the opening of the superconducting gap. This enhancement also depends upon the coherence factor and as seen for the $s^{+-}$ case, if the coherence factor vanishes, the onset of the damping rises slowly as $\omega$ exceeds $2\Delta_0$ and is smaller than in the normal state.

In Fig.~\ref{fig:3}b, we have plotted the difference between the real part of the phonon self-energy in the superconducting and the normal states asociated with the $xz$--$yz$ scattering processes. The $s^{++}$ case shows the familiar hardening of the phonon that occurs when $\Omega_{Q_1}$ exceeds $2\Delta_0$. That is, in the $s^{++}$ superconducting state, ${\rm Re}\delta\Pi(Q_1,\omega)$ is positive for $\omega>2\Delta_0$. As we will see when we look at the phonon spectral weight, the negative dip for $\omega\ltwid2\Delta_0$ leads to what Allen et al.\cite{ref:Allen} have called a ``vibrational/superelectronic" resonance. The structure of ${\rm Re}\delta\Pi(Q_1,\omega)$ is quite difference in the $s^{+-}$ state. Here for $\omega>2\Delta_0$, $\delta{\rm Re}\Pi(Q_1,\omega)$ is negative and this leads to a softening of the $\Omega_{Q_1}$ phonon. In addition, there will not be a resonance feature below $2\Delta_0$ in the phonon spectral weight. 

\begin{figure}
	[htbp] 
	\includegraphics[width=8.5cm]{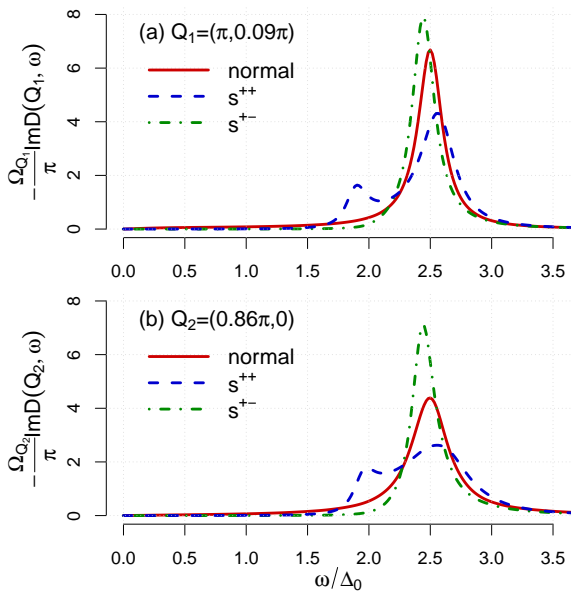} \caption{(Color online) The phonon spectral weight $-\frac{\Omega_{Q}}{\pi}{\rm Im} D(Q,\omega)$  versus $\omega/\Delta_0$ for wavevectors (a) $Q_1$ and (b) $Q_2$ for the normal state (red, solid), the $s^{++}$ (blue, dashed) and $s^{+-}$ (green, dot-dashed) superconducting gaps. Note that we have added a finite damping in the calculation of the orbital susceptibility $\chi^{\rm orb}(q,\omega)$. \label{fig:4}} 
\end{figure}

Using the results for the self-energy shown in Fig.~\ref{fig:3}, we have plotted the phonon spectral weight $-\frac{\Omega_{Q_1}}{\pi}{\rm ImD}(Q_1,\omega)$ versus $\omega/\Delta_0$ in Fig.~\ref{fig:4}a. The solid line shows the result for the normal state while the long dashed and the dash-dot curves are for the superconducting $s^{++}$ and $s^{+-}$ states, respectively. Here one sees that for an $s^{++}$ gap there is the expected resonance just below $2\Delta_0$ and the phonon peak has shifted up in frequency and broadened. For the $s^{+-}$ gap, the anomalous resonance below $2\Delta_0$ is absent and the phonon peak has slightly softened as well as narrowed. Similar results for the $Q_2$ phonon spectral weight are shown in Fig.~\ref{fig:4}b.

In RPA calculations \cite{ref:Kuroki2,ref:Graser,ref:Kemper} there were parameter ranges in which the $s^{+-}$-like gap was found to have nodes on the $\beta$ Fermi surfaces. In this case, the nominally $s^{+-}$ gap will have the same sign on the parts of the Fermi surfaces connected by $Q_2$. However, they would differ for $Q_1$. Thus, it will be important to measure the phonon spectral weight for various near nesting $Q$ values. However, one will need phonon modes that couple to the relevant orbital scattering and in addition have frequencies $\Omega_Q$ of order the sum of the magnitudes of the superconducting gaps $|\Delta(k+Q)|+|\Delta(k)|$ associated with the nearby nested regions. In principle, with sufficient $q$ and $\omega$ resolution and sensitivity, neutron and x-ray scattering offer the possibility of mapping out the relative structure of the gap between various parts of the Fermi surfaces in Fe-based as well as other unconventional superconductors.


\section*{Acknowledgements}

We would like to acknowledge useful discussions with T.~Devereaux and A.F.~Kemper. PH and DJS would like to thank the Stanford Institute for Materials and Energy Sciences for their hospitality. TAM and DJS would like to acknowledge support from the Center for Nanophase Materials Sciences, which is sponsored at Oak Ridge National Laboratory by the Office of Basic Energy Sciences, U.S. Department of Energy. This work was supported by DOE DE-FG02-05ER46236 (PJH), the DFG through TRR80 and the Free State of Bavaria through the BaCaTeC program (SG).

\end{document}